\documentstyle[11pt]{article}
\titlepage
\newcommand{\be}{\begin{equation}}

\newcommand{\ee}{\end{equation}}

\newcommand{\bea}{\begin{eqnarray}}
\newcommand{\eea}{\end{eqnarray}}

\begin{document}


\title
{ On the Cosmological Origin of the Homogeneous Scalar Field \\
in Unified Theories}

\author{
V.N.Pervushin, V.I Smirichinski \\[0.3cm]
{\normalsize\it Joint Institute for Nuclear Research},\\
 {\normalsize\it 141980, Dubna, Russia.}}

\date{\empty}

\maketitle
\medskip


\begin{abstract}
{\large
{ We consider the possibility of describing the Higgs effect in
unified theories without the Higgs potential in the presence of the
Einstein gravity with the conformal gravity-scalar coupling under
the assumption of homogeneous matter distribution.

  The scalar field values can be found from the Friedmann equations for
the homogeneous Universe. The considered cosmological mechanism solves the
vacuum density problem (we got $\rho_\phi^{Cosmic}=10^{-34}\rho_{cr}$
instead of $\rho_\phi^{Higgs}=10^{54}\rho_{cr}$),
and exludes the monopole creation and the domain
walls.
}}

\end{abstract}

\newpage

\section{Introduction}
  The homogeneous scalar field, generating elementary particle masses in
unified theories, is based on the Higgs potential.
The physical motivation for this potential as a consequence
of the first symmetry principles is unclear and the existence of this
potential leads to a number of difficulties in cosmology. Among them are the
 great
vacuum density ~\cite{1}, monopole creation ~\cite{2}, domain walls ~\cite{3}.
 These difficulties are overcome in the inflationary models based on the
assumption of minimal scalar - gravity coupling ~\cite{4}.
  The Higgs potential, as well as the minimal coupling, break the principle
about conformal invariance of gravity - matter interaction ~\cite{5}.
  If we take into consideration this conformal symmetry principle,  it is
necessary to switch off the Higgs potential and to introduce a conformally
invariant scalar field-gravity interaction ~\cite{5}. As a result, the only
source of conformal symmetry breaking is the Einstein gravity itself.

In the present paper, we shall show that the
Einstein gravity theory with a conformally coupled scalar field
 can replace the Higgs potential
 under the assumption of homogeneous matter distribution.
 We shall also find the
scalar field value from the Friedmann equations for the homogeneous Universe
with a Friedmann-Robertson-Walker metric (FRW).

\section{The Higgs effect without the Higgs potential in the FRW metric}

 We begin from the $U(1)$ theory with the Lagrangian
\be     \label{L1}
L(A,\phi) = |(\partial_{\mu}+ieA_{\mu})\phi|^2 -
\frac{1}{4}F_{\mu\nu}(A)F^{\mu\nu}(A) - V_{Higgs}(|\phi|)
\ee
with a complex scalar field $\phi =|\phi| \exp (ie\chi)$.
The Higgs effect consists in i) the absorption of the angular component
of the scalar field by the transformation
\be     \label{T}
  A'_{\mu}=\partial_{\mu}\chi + A_{\mu} \\
\ee
\be     \label{L2}
L(A',\phi) = \partial_{\mu}|\phi|\partial^{\mu}|\phi| +
e^2 |\phi|^2 A'_{\mu}A'^{\mu} -
\frac{1}{4}F_{\mu\nu}(A')F^{\mu\nu}(A') - V_{Higgs}(|\phi|)
\ee
and in ii) the nonzero vacuum value of the scalar field module which
follows from the Higgs potential.

For computing this vacuum value one commonly minimizes the Lagrangian
(\ref{L2}) without interaction
\be     \label{L3}
L_{o}(A',\phi) = \partial_{\mu}|\phi|\partial^{\mu}|\phi| -
\frac{1}{4}F_{\mu\nu}(A')F^{\mu\nu}(A') - V_{Higgs}(|\phi|).
\ee
Our idea is to consider just this theory (~\ref{L3})
without the Higgs potential ($V_{Higgs}(|\phi|)=0$) but in the presence
of the Einstein gravity theory with a conformally coupled scalar field
\be     \label{L4}
L_{o}^{G}(g,A',|\phi|) = \sqrt{-g}\left[ -\frac{^{(4)}R(g)}{6}
(\frac{3M_{Pl} ^2}{8\pi}-\left| \phi \right| ^2)+\partial_\mu \left| \phi
\right| \partial ^\mu \left| \phi \right| -
\frac{1}{4}F_{\mu\nu}(A')F^{\mu\nu}(A')\right].
\ee
To find  minimum of the action with the Lagrangian (\ref{L4}), we
express it in terms of the conformal invariant variables, marked by (c),
extracting the space-scale factor \cite{6} \\

\be     \label{a}
a= [{}^{(3)}g]^{1/6}.
\ee

\be     \label{C1}
g_{\mu\nu}= a^2 g^{(c)}_{\mu\nu}~; A_{\mu}=A^c_{\mu};~~
\left| \phi \right| =\frac{ \phi_c }{a},
\ee
where the space components of the new metric $g_{(c)}$ satisfy the
 constraint
$\sqrt{{}^{(3)}g^{(c)}}=1$ by the definitions (\ref{a}) and (\ref{C1}).
 Lagrangian (~\ref{L4}) has the symmetric form  with respect to
$a$ and $ \phi_c $ with ${\sqrt{-g^{(c)}}} = N_c$:
\be     \label{L7}
L_{o}^{G} = N_c\left[ -\frac{^{(4)}R(g^{(c)})}{6}
(\frac{3M_{Pl} ^2}{8\pi}a^2- \phi_c  ^2)+
\partial_\mu  \phi_c  \partial^\mu  \phi_c  -\\
\frac{3M_{Pl} ^2}{8\pi}\partial_\mu a \partial^\mu a
- \frac{1}{4}F_{\mu\nu}(A')F^{\mu\nu}(A')\right].
\ee
The supposition about the homogeneous distribution of the field $A'$ can
lead to homogeneity of both the scale factor $a$ and scalar field
$ \phi_c $ as the motion equation for scalar field
repeats the one for the scale factor.

To get the homogeneous sector of the theory (\ref{L7}), it is sufficient to
use the isotropic version of the FRW metric
\be  \label{FRW1}
(ds)^2=a^2(x,t)[N_c^2(x,t)dt^2-dx_i^2],
\ee
\be  \label{FRW2}
a(x,t)=a_0(t)S(x)\;;\;\;N_c(x,t)=\frac{N_c^0(t)}{S(x)}\;;\;\;
S=(1+\frac{k|x|^2}{4r_0^2})^{-1};~~~~(k = +1; 0; -1).
\ee
In order to calculate the homogeneous scalar field $\phi_c$ we restrict
ourselves to the case of constant positive space curvature (1)
$k=+1$ with a volume $V_{(3)}=\int d^3xS^3(x)=2\pi^2r_0^3$
and consider harmonic excitations of the vector field $A'$, in this space,
with harmonics $\omega_l$ described by the action
\be     \label{LA}
W_{R}^{H}(A) =\int dtd^3x N_c\left[
- \frac{1}{4}F_{\mu\nu}(A')F^{\mu\nu}(A')\right]=
\frac{V_{(3)}}{2}\int_0^{t_1} dt
\sum_l \left[\frac{dA_l^2}{N_c^0{dt}^2} -
N_c^0\omega_l^2A_l^2\right]
\ee
The substitution of (\ref{FRW2}) and (\ref{LA}) into the Lagrangian (\ref{L7})
leads to the action
\be \label{WH}
W^H=V_{(3)}\int_0^{t_1} dt\left[-\frac{3M_{Pl}^2}{8\pi}\left(
\frac{da{}_0^2}{N_c^0dt^2}-N_c^0\frac{a_0^2}{r_0^2}\right)+
\left(\frac{d\phi_{c}^{2}}{N_c^0dt^2}-
N_c^0\frac{\phi_c^2}{r_0^2}\right)\right] + W_R^H.
\ee
This action describes a set of  oscillators evaluated in respect to
the invariant conformal time $d\eta=N_c^0dt$ (see Appendix) with
conserved energy densities
\be \label{rho}
\rho_{cr}=\frac{3M_{Pl}^2}{8\pi}\left(
\frac{da{}_0^2}{d\eta^2}+\frac{a_0^2}{r_0^2}\right);~~
\rho_{\phi}^0=\left(\frac{d\phi_{c}^{2}}{d\eta^2}+
\frac{\phi_c^2}{r_0^2}\right);~~
\rho_{R}=\sum_l \left(\frac{dA_l^2}{{d\eta}^2} +
\omega_l^2A_l^2\right)
\ee
connected by the constraint (Einstein - Friedmann equation
$ \delta W^H / \delta N_c^0=0 $)
\be \label{rho3}
   - \rho_{cr} + \rho_{\phi}^0 + \rho_{R} = 0.
\ee

The geometrical observables of the Friedmann Universe, in comoving frame of
reference, are constructed
by using the inverse conformal transformation (\ref{C1}) of the dynamical
variables and coordinates,
including the Friedmann time interval $dt_F=a_0 d\eta$ and distance
 $D_F=a_0 D_c$ \cite{6,7}.
The evolution of the cosmic scale $a(t_F)$, in the considered case, coincides
with the one of the Friedmann Universe filled by radiation
\be
t_F^{(\eta)} = r_0\Omega_0^{-\frac{1}{2}}(1-\cos\frac{\eta}{r_0})\;\;\;; \;\;\;
\Omega_0 = \frac{3M_{Pl}^2}{8\pi r_0^2 \rho_{cr}}
\ee
\be  \label{c16}
a(\eta)= \Omega_0^{-\frac{1}{2}} \sin\frac{\eta}{r_0}=
\left[\frac{t_F}{r_0}(2\Omega_0^{\frac{1}{2}}- \frac{t_F}{r_0})\right]^{1/2}
\ee
\be    \label{e17}
\eta =r_0 \; arccos[1-\Omega_0^{\frac{1}{2}}t_F/r_0],
\ee
with the Hubble constant
\be
\left.
H_0=\frac{da}{adt_F}\equiv\frac{d\phi_c}{\phi_cdt_F}=\frac{1}{t_F}
\left[\frac{1-(t_F/r_0)\Omega_0^{1/2}}{2-(t_F/r_0)\Omega_0^{1/2}}.
\right]
\right|_{r_0\to\infty}\simeq\frac{1}{2t_F}\;.
\ee
The scalar field $\phi_c$ repeats  this evolution (\ref{c16})
\be
\phi_c(t_F)= M_{Pl}\left(\frac{3\rho_\phi^0}{8\pi\rho_{cr}}
\right)^{\frac{1}{2}}a(t_F).
\ee
While the initial scalar field $|\phi|$ defined by eq.(7) is equal to a constant
\be
|\phi|=\frac{\phi_c}{a}= M_{Pl}\left(\frac{3\rho_\phi^0}{8\pi\rho_{cr}}
\right)^{\frac{1}{2}}.
\ee
The value of this scalar-field, which follows from the Weinberg-Salam theory
$|\phi| \sim 10^2$GeV, allows us to estimate the value of the relation of energy
densities of the scalar field $(\rho_{\phi}^0)$ and the expansion of the
Universe
$(\rho_{cr})$:
\be
\rho_\phi^{Cosmic}=10^{-34}\rho_{cr}
\ee
Recall, that the Higgs potential leads to the opposite situation
(see \cite{1})
\be
\rho_\phi^{Higgs}=10^{54}\rho_{cr}.
\ee
 The homogeneity of the scalar field (as the consequence of the homogeneous
 distribution of matter) excludes monopoles \cite{2} and domain walls
 \cite{3}.


\section{ Weinberg-Salam model cosmology without Higgs potential}
We consider the Einstein theory supplemented by the conformal invariant
part of the Weinberg-Salam theory
\be     \label{L44}
W_{tot} =\int dtdx^3 \sqrt{-g}\left[ -\frac{^{(4)}R(g)}{6}
(\frac{3M_{Pl} ^2}{8\pi}-\left| \phi \right| ^2)+\partial_\mu \left| \phi
\right| \partial ^\mu \left| \phi \right| \right] + W_{WS}^c(\{{}^{(n)}F\}).
\ee
 where $W_{WS}^c$ depends on the entirety of fields $\{ {}^{(n)}F\}$             with conformal
weights $(n)$:
gravitational  $(g)\;n_g=2;$ , vector $(A,B)\;\;n_{A,B}=0;$
 spinor  $(e,\nu)\;\;n_{\nu,e}=-\frac{3}{2};$
 the doublet of
the scalar field $(\phi)\;n_{\phi}=-1$ with a module $|\phi|$

$$
\phi=\left|\begin{array}{c} \phi_1 \\ \phi_2 \end{array}\right|=
|\phi|\left(\begin{array}{c} n_1 \\ n_2 \end{array}\right);\;\;
n_1\stackrel{*}{n_1}+n_2\stackrel{*}{n_2}=1,
$$
the angular components of wich ($n_1,n_2$) are absorbed the vector
field ($W,Z$):
\be
\begin{array}{r}
W^c_{WS}=\int d^4x\sqrt{-g} \left[ |\phi|^2 \left( W_\mu W_\nu\frac{g^2}{2}+
Z_\mu Z_\nu \frac{g^2+g_1^2}{2} \right) g^{\mu \nu}- |\phi| \frac{G_e}{2} \bar e e \right.\\ \\
+ i\bar\nu{}_e\hat\partial{}_{Fock}\nu_e+i\bar e{}\hat\partial{}_{Fock}e-
\frac{1}{2}G_{\mu\nu}^a(A^a)G^{a\mu\nu}(A^a)-\frac{1}{4}F_{\mu\nu}(B)F^{\mu\nu}(B)\\ \\
+\frac{gg_1}{(g^2+g_1^2)^{1/2}}\bar e\gamma^\mu e A^\mu-
\frac{g}{2\sqrt{2}}[\bar\nu{}_e\gamma^\mu(1+\gamma^5 )e W_\mu
+\bar e \gamma^\mu(1+\gamma^5)\nu W^*_\mu]\\ \\ \left.
+\frac{(g^2+g_1^2)^{1/2}}{4}[\bar\nu{}_e\gamma^\mu(1+\gamma^5)\nu_e-2
\bar e\gamma^\mu \left( \gamma^5+\frac{g^2-3g_1^2}{g^2+g_1^2} \right)e]Z_\mu
\right],
\end{array}
\ee
\be
Z_\mu=(g^3+g_1^2)^{-1/2}(-gA^3_\mu+g_1B_\mu)\;\;;\;\;
A_\mu= (g^2+g_1^2)^{-1/2}(g_1A^2_\mu+gB_\mu)
\ee
\be
W_\mu(\pm)=\frac{A_\mu^1\pm A_\mu^2}{\sqrt{2}};
\ee
where ($\hat\partial{}_{Fock}$) the Fock covariant derivation in metric
$g_{\mu\nu}$, $G^a_{\mu\nu}$ is the Yang-Mills tension for $A_\mu^a$, $F_{\mu\nu}$ is
$U(1)$-tension for $B_\mu$.

 As we have seen in the previous section,
 the central point in the derivation of the dynamics of the cosmological
model from field theory is the relation between the geometric observables in
the Einstein theory (\ref{L44})$\{ {}^{(n)}F\}$ and the dynamical observables
 $\{ {}^{(n)}F_c \}$
of the Lagrangian (or Hamiltonian) approach to the cosmological model
 of the expanding Universe
\be
{}^{(n)}F_á={}^{(n)}F_c a^{+n},
\ee
 for which the integrals of motion are found ~\cite{6,7}. This relation can be
formulated
in the form of a principle about the conformal invariance of dynamical
variables $\{ {}^{(n)}F_c \}$.
  In the same way the dust mass in the Friedmann Universe $M_D$ appears
from the electrodynamical action
\be \label{D}
M_D=m\int d^3x\sqrt{g^{(3)}}\bar\psi\psi=m\int d^3xa^3\bar\psi\psi
=m<n_f>V_{(3)},
\ee
as an integral of motion   ($\dot M{}_D=0$),
if $\psi=a^{-3/2}\psi_c$.

Due to the conformal invariance the action $W_{WS}^c$  does not depend on the
scale variable ($a$).

Let us calculate the homogeneous scalar field in this theory (\ref{L44}) in
the supposition about the homogeneous distribution of all matter fields in
the Universe. The action $W_{WS}^c$ does not change the evolution of the cosmic
scale factor (\ref{c16}) and can lead only to the additional terms, in the energy density
of the scalar field of the type of (\ref{D}),
\be
 \rho_\phi=\rho_\phi^0-\phi_c<n_f> +\phi_c^2 <n_b^2>,
\ee
associated with the fermion and boson "dusts" at rest, the masses of which are
formed by the homogeneous scalar field itself. Here $\rho_\phi,<n_f>,<n_b^2>$
are phenomenological parameters which determine the solution to the
 homogeneous
scalar field equation. For the  case considered we have

\be
\phi_c(\eta) = \rho_\phi^{1/2}\sin\omega_\phi\eta+\frac{1}{2}<n_f>\omega_\phi^{-2}
(1-\cos \omega_\phi\eta),
\ee
where $\eta$ is defined by eq.(\ref{e17}), $\omega^2_\phi=1/r_0^2+<n_b^2>$.

If the dust term dominates and $\omega_\phi\not= 1/r_0$,
the WS-particle masses ($\phi_c/a$) become  dependent on time.
A photon radiated by an atom on an astronomical object (with a distance $D$
to the Earth) at the time $t_F -D$ remembers the value of this mass at
this time. As the result, the red shift and the Hubble law, in the comoving
frame of reference, is defined by the product of two
factors: the expansion of the Universe space $(a)$  and the alteration of the
elementary particle masses $(\phi_c/a)$
\be
a\left(\frac{\phi_c}{a}\right)=\phi_c.
\ee
Finally, for the unified theory version of the homogeneous Universe we got
the
red shift $Z$ and the Hubble law
\be
Z(D)=\frac{\phi_c(t_F)}{\phi_c(t_F-D)}-1\;\;;\;\;H_0=\frac{d\phi}{\phi_cdt_F}.
\ee
If $<n_b^2>=0$ and dust dominates, $H_0=t_F^{-1}$. If $<n_b^2>\gg 1/r_0^2 $ we got the
oscillator-like behavior of the red shift which can immitate the large scale
structure of the Universe \cite{10}.

\section{Conclusion}
  We tried to describe a homogeneous scalar field in the Weinberg-Salam
theory,
unified with Einstein gravity, starting from the assumption that the only
source of conformal invariance breaking (on a classical level) is the Einstein
gravity. This means that the investigated version of unified theory does not
contain the Higgs potential and is based on the conformal invariant scalar
field - gravity coupling. In such a theory the homogeneous scalar field is
calculated from the Friedmann equations for the homogeneous Universe. As a
result, one of the versions of the physical realizations of Max's principle
appears, namely that the mass of elementary particle forming the matter
are determined by the distribution of this matter in the Universe.

 Some new consequences of the investigated version of the unified theory are
a somewhat different
 Hubble law for the dust Universe and the cosmological evolution of
elementary particle masses in the comoving reference frame. Such a mass
evolution can lead to a gradual (in a cosmological scale) decrease of
the relative distance between the gravitating objects.
The principle of conformal invariant interaction of matter fields with
the Einstein gravity and the conventional Friedmann assumption of
homogeneous matter distribution
lead to very small energy density of scalar field and exclude the monopole
creation and domain walls.

\vspace{0.5cm}

Acknowledgments.

We are happy to acknowledge interesting and critical
discussions with Profs. B. Dimitrov, A.V.Efremov, A.T. Filippov, A.M. Khvedelidze,
V.A. Rubakov, A.A. Slavnov, E.A.Tagirov, I.T. Todorov, and J. Zinn-Justin
and to thank the Russian Foundation
for Basic Investigation, Grant N 96\--01\--01223, for support.

\newpage

\appendix
\begin{center}
{\bf Appendix}
\end{center}
\setcounter{equation}{0}
\renewcommand{\theequation}{A.\arabic{equation}}
The  actions of homogeneous models considered here can be represented in the
Hamiltonian form
\be \label{6}
W^H(p_a,a,F)=\int_0^{t_1} dt\left[-P_a\frac{da}{dt} + \sum_F P_F \frac{dF}{dt}
-N_c^0 (-H_a + H_F)\right]
\ee
where
\be \label{ci}
H_a = \frac{P_a^2}{V_{(P)}} + \frac{a_0^2}{r_0^2}V_{(P)};~~~V_{(P)}= V_{(3)}
\frac{3M_{Pl}^2}{8\pi},
\ee
$H_F$ depends only on a set of the field harmonic excitations in the
FRW metric space, and it does not depend on the cosmic scale $a$.

The reduction of the constrained (\ref{6}) to an equivalent unconstrained
one can be fulfilled by the
canonical transformation  to the new cosmic variables ~\cite{6,7,11}:
\be
(P_a, a)\;\Longrightarrow\;  (\Pi, \eta)\;\;;\;\;\{P_a,a\}_{(\Pi,\eta)}=1;~~~
\{ P_a,a\}_{(\Pi,\eta)}=1,
\ee
so that the cosmic part of the constraint $-H_a+H_F=0$ converts into a new
momentum $\Pi$: $H_a=\Pi$. This equation represents a map of the circle
(\ref{ci}), in the old phase space, into a line, in the  new phase space.
There are two  maps of this type
\be  \label{8}
P_a=\pm\sqrt{V_{(P)}\Pi}\cos\frac{\eta}{r_0}\;;\;
a_0=\pm\sqrt{V_{(P)}^{-1}\Pi}\sin\frac{\eta}{r_0};
\ee
Thus, we got two actions
 instead of (\ref{6}):
\be  \label{9}
W_{\pm}^H=\int_0^{t_1} dt \left[\pm \Pi \frac{d\eta}{dt}+\sum_{F}
\frac{dF}{dt}-N_c^0 (-\Pi + H_F))\right].
\ee
In this version of the theory the equations of motion for $\Pi$
\be
\frac{\delta W_\pm}{\delta \Pi}=0\;\Longrightarrow\;\pm d\eta=N^0_cdt
\ee
determine the invariant parameter of the dynamical evolution of the variables
in the theory in the sector of the Dirac observables $F$  and
the equation for $N_c^0$ gives the Hamiltonian for such an evolution.

\be
\frac{\delta W_\pm}{\delta N_c}=0\;\Longrightarrow\;\Pi=H_F
\ee
As a result we receive the equivalent (\ref{9}) dynamical system without
constraints
\be  \label{12}
W_{\pm}^{Red}=\int_0^{\eta(t)}
 d\eta\left[\sum_{F}P_F\frac{dF}{d\eta}\mp H_F\right].
\ee
In this approach, we can see that the new cosmic variable becomes invariant
time of evolution in the sector of the Dirac observables.

\end{document}